\documentclass[article]{aa}  

\usepackage{natbib}
\bibpunct{(}{)}{;}{a}{}{,} 
\usepackage{graphicx}
\usepackage{txfonts}

\newcommand{\dx}{\mathrm{d}x}
\newcommand{\dy}{\mathrm{d}y}

\usepackage{hyperref}
\usepackage{multirow}

\begin{document}

   \title{Evaluating the effect of stellar multiplicity on the PSF \\
   of space-based weak lensing surveys}

   \subtitle{}

   \author{T. Kuntzer
          \inst{1}
          \and
          F. Courbin\inst{1}
          \and
          G. Meylan\inst{1}
          }

   \institute{Laboratoire d'astrophysique, Ecole Polytechnique F\'ed\'erale de Lausanne (EPFL), Observatoire de Sauverny, CH-1290 Versoix, Switzerland 
            }

   \date{Received month day, year; accepted month day, year}

  \abstract
    {The next generation of space-based telescopes used for weak lensing surveys will require exquisite point spread function (PSF) determination. Previously negligible effects may become important in the reconstruction of the PSF, in part because of the improved spatial resolution. 
    In this paper, we show that unresolved multiple star systems can affect the ellipticity and  size of the PSF and that this effect is not cancelled even when using many stars in the reconstruction process. We estimate the error in the reconstruction of the PSF due to the binaries in the star sample both analytically and with image simulations for different PSFs and stellar populations. 
    The simulations support our analytical finding that the error on the size of the PSF is a function of the multiple stars distribution and of the intrinsic value of the size of the PSF, i.e. if all stars were single. Similarly, the modification of each of the complex ellipticity components $(e_1, e_2)$ depends on the distribution of multiple stars and on the intrinsic complex ellipticity. 
    Using image simulations, we also show that the predicted error in the PSF shape is a theoretical limit that can be reached only if large number of stars (up to thousands) are used together to build the PSF at any desired spatial position. 
    For a lower number of stars, the PSF reconstruction is worse. Finally, we compute the effect of binarity for different stellar magnitudes and show that bright stars alter  the PSF size and ellipticity more than faint stars. This may affect the design of PSF calibration strategies and the choice of the related calibration fields.
    }


   \maketitle
%

\section{Introduction}
\defcitealias{Duchene2013}{DK13}

Weak gravitational lensing is one of the main cosmological probes to study dark energy and dark matter \citep[e.g.][]{Weinberg2013, Frieman2008}. In practice, this statistical method requires measuring the shape of billions of faint and small galaxies with the least possible contamination by the instrumental point spread function (PSF). Space-based wide field surveys provide the highest quality check to perform these challenging measurements. Two missions are either under construction or in project to conduct these wide field optical and near-IR surveys: the ESA Euclid satellite\footnote{\url{http://www.euclid-ec.org/}} \citep{Euclid} to be launched in 2020 and the NASA Wide Field InfraRed Survey Telescope \citep[WFIRST; e.g.][]{WFIRST}.

Even with the high spatial resolution of a space telescope, any weak lensing experiment must account for the small but non-negligible instrumental PSF, which needs to be removed accurately.
The weak lensing distortions are an order of magnitude smaller than the PSF effects on the shape of the observed galaxies \citep[e.g.][]{Kilbinger2015}. 
To achieve the required accuracy, the PSF must be characterised using stars in the field of view. However, in practice, the high spatial reso\-lution needed to resolve faint galaxies also means that double or multiple stars may significantly affect the PSF measurement, i.e. its size and ellipticity. 
The goal of the present work is to assess the bias and error introduced by stellar multiplicity on the PSF determination, which is crucial for weak lensing experiments \citep[e.g.][]{Cropper2013, Massey2013}.

Binary and multiple stellar systems are very common \citep[see review by][hereafter DK13]{Duchene2013}; they formed in and their cha\-racteristics depend on the initial environment and mass \citep{Reipurth2014}. 
Several volume and magnitude-limited surveys were dedicated to the search of binary stars, allowing a rigorous construction of catalogues with reliable distribution inferences beginning with the work of \citet{Duquennoy1991} and compiled by \citetalias{Duchene2013}. Nevertheless, large uncertainties remain in the distribution of the multiple system parameters, even in the solar neighbourhood, reflecting the difficulty of such surveys in accounting for the properties of multiple stellar systems in detail. 
The frequency of multiple system varies drastically depending on the mass of the primary object, with almost 100\% of the most massive stars being multiple systems and on average 35--40\% of main-sequence stars being at least binary \citepalias{Duchene2013}. 

\begin{figure*}
 \centering
  \includegraphics[width=0.23\linewidth]{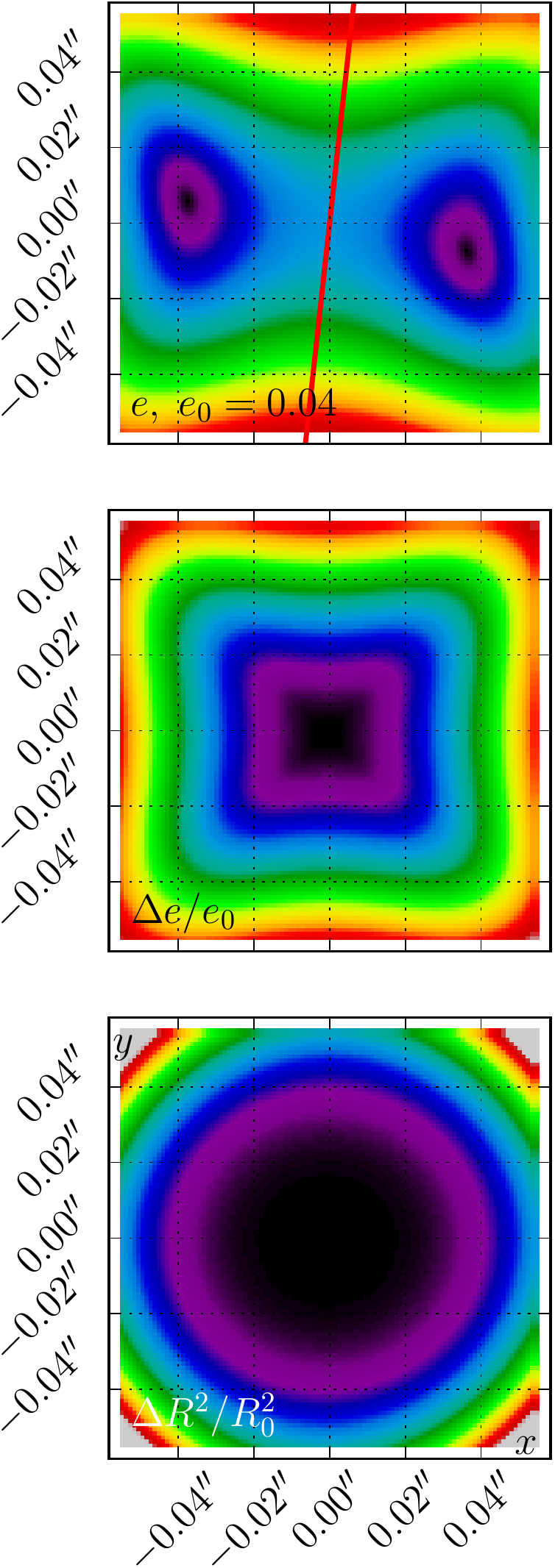}
  \includegraphics[width=0.253\linewidth]{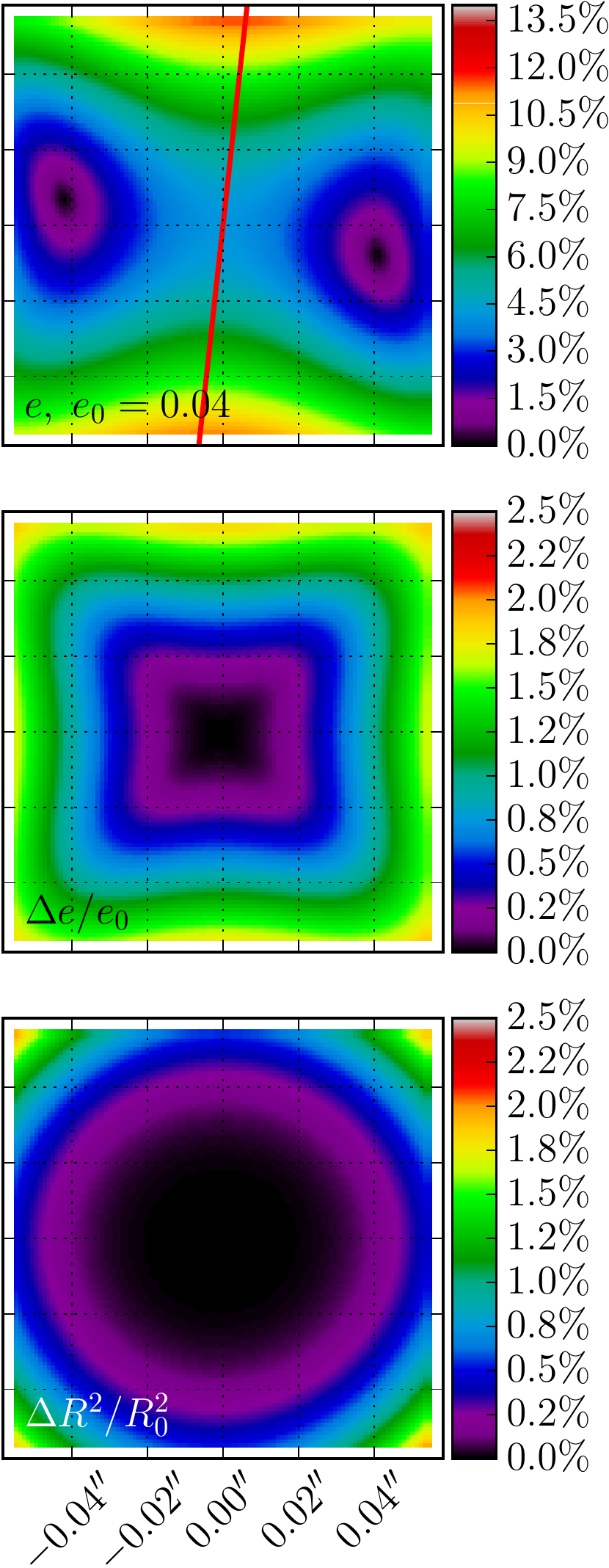}
\hfill
  \includegraphics[width=0.23\linewidth]{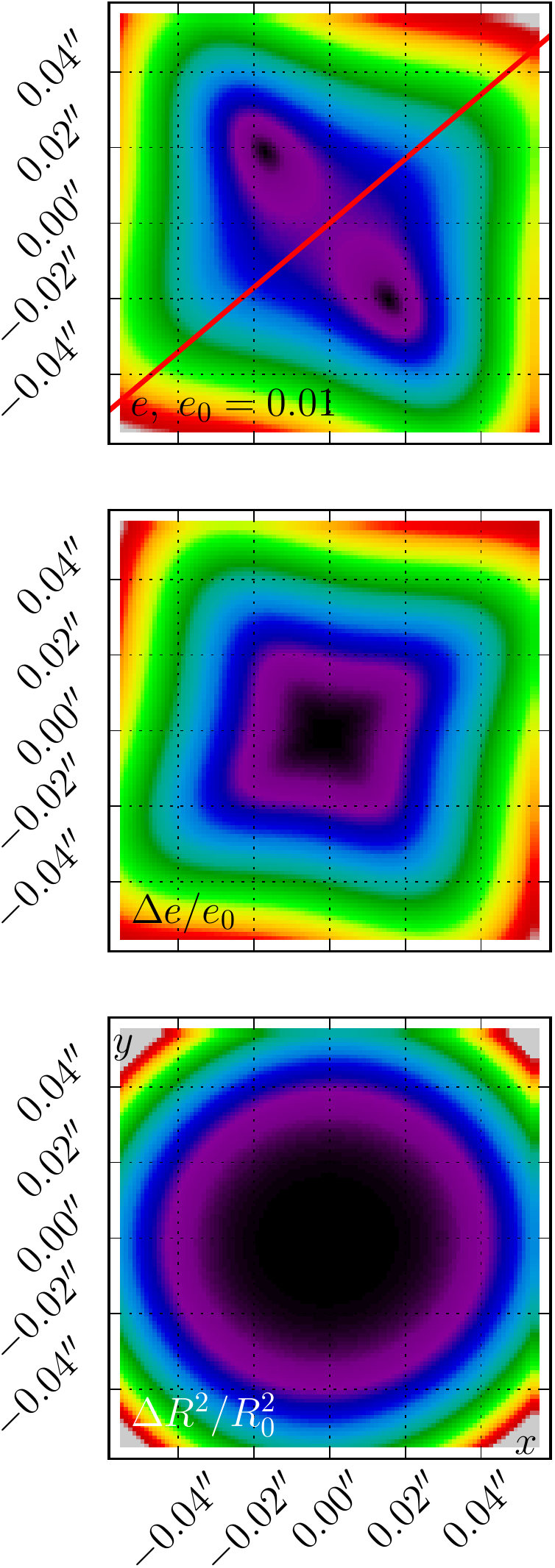}
  \includegraphics[width=0.253\linewidth]{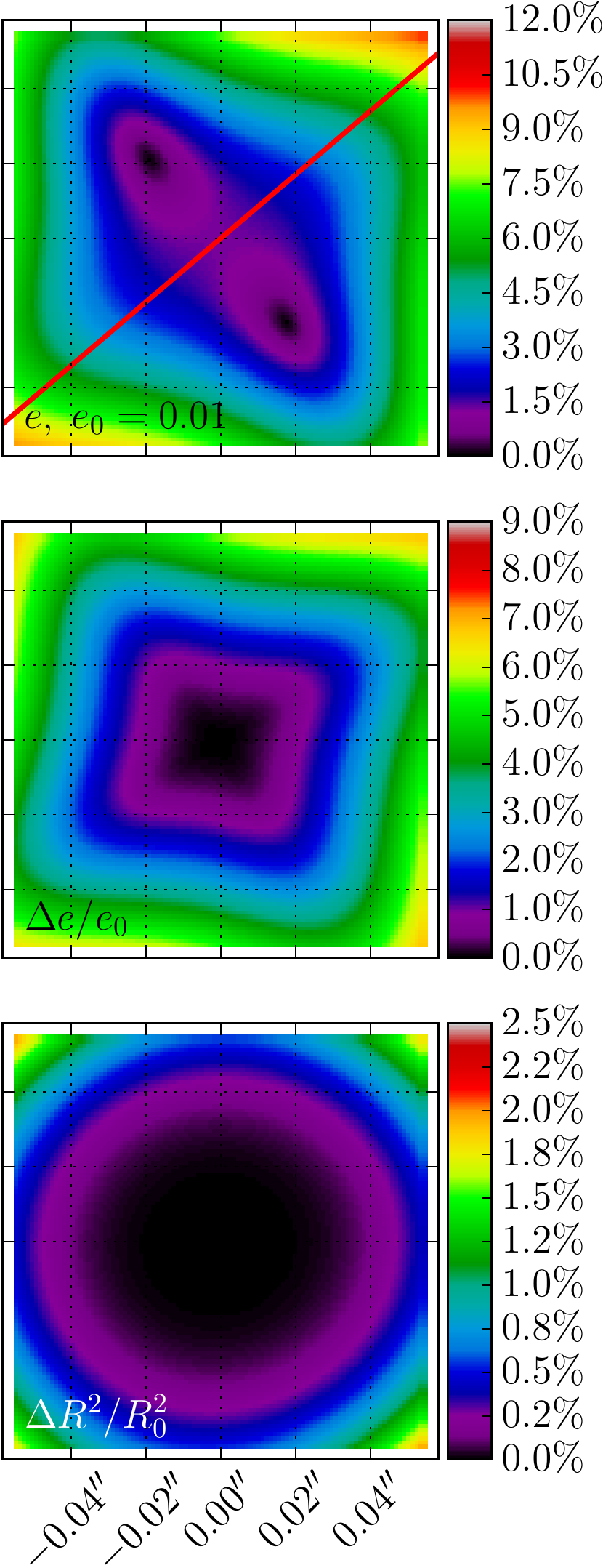}
  \caption{\label{fig:single}
  Reconstruction errors for a single binary star with the main component centred on (0,0). Each point on the map shows the measurement of the quantity labelled in each panel when a single companion star is located at that position. The labels are in arcsecond. 
  The first column on the left shows the effect for a PSF aligned along $e_1$ and a contrast of $\Delta m =0$ between the main star and its companion. The next column is for the same PSF shape and orientation, but for a contrast of $\Delta m =1$. 
  The block of six graphs on the right is for the same two contrasts, but for a PSF aligned along $e_2$.
  The red line represents the orientation of the PSF.
 From top to bottom, the total ellipticity (star + companion) of the PSF, the relative error on the ellipticity due to the companion star, and the relative error on the size are shown.}
\end{figure*}

In this paper, we simulate the error in the reconstruction of the PSF arising from stellar multiplicity. The information on the distribution of multiple systems and their properties (separation, contrast) is fairly scarce but remains sufficient to simulate a space-based survey. 
For this purpose, we use the Besan\c{c}on model of the Galaxy \citep[BMG;][]{Robin2004} in combination with the binary star fraction characteristics drawn from observed stellar distributions summarised in \citetalias{Duchene2013}. 
These data allow us, first, to build mock catalogues with realistic stellar properties and, second, to simulate space-like images of stars. 
With these simulations, mocks, and images, we derive the bias and reconstruction errors on the PSF width and ellipticity. We also derive analytical predictions and show that the analytical predictions agree very well with the results of the simulations, even for a realistic space-like PSF with complex structures and spikes.

We first study the effect of one single companion as a toy model of the problem in Sect.~\ref{sec:toymodel}.
In Sect.~\ref{sec:th}, a formalism is developed to analytically compute the expected PSF reconstruction error and bias.
In Sect.~\ref{sec:binaries}, we discuss the current knowledge of the parameters of multiple systems in the Milky Way,  describe how we construct the stellar population, and  detail the simulations that support our analytical findings. 
We present the results of the simulations and their parameters in Sect.~\ref{sec:discussion} and conclusions are drawn in Sect.~\ref{sec:conclusions}.

\section{Reconstruction error of PSF for a single binary star} 
\label{sec:toymodel}

The typical physical separations in binary stars are of the order $a=50$~AU and typical light contrast is between 0 and 1 magnitude. With the resolving power of a Euclid-like telescope, 0.1\arcsec, binary stars can be resolved up to distances of $d=2.5$~kpc. 
These resolved binaries can be identified and removed from samples of stars to be used for the PSF reconstruction. However, most binaries cannot be explicitly identified and still affect the PSF determination, which is an effect that we illustrate in this section.

We generate a simple simulated image that uses a space-like PSF with the {\tt GalSim} software \citep{Rowe2015}. This simulation is composed of a main star located in the centre of the image to which we add a companion star at a given position within the central pixel. 
We then measure the error in the reconstruction of the PSF of the double star and compare its ellipticity and size to the values found for a single star.

The resulting errors are presented in Fig.~\ref{fig:single} for two different ellipticities and for two different contrasts in magnitude, $\Delta m =0$ and $\Delta m =1$. Each PSF has its major axis along $e_1$ and $e_2$, respectively. The colour maps in Fig.~\ref{fig:single} show the ellipticity of the reconstructed PSF, the relative errors on the ellipticity, $\Delta e/e_0$, and on the size, $\Delta R^2/R^2_0$, where $e_0$ and $R_0^2$ are the intrinsic ellipticity and size of the PSF in the absence of a companion. 

Two effects are striking. First, the relative errors can easily reach 1\% in ellipticity when the companion star is as close as 0.01\arcsec\ from its host. This figure is reached when the contrast in magnitude is $\Delta m=0$. Even for a larger contrast of $\Delta m=1$, the error remains of the same order of magnitude. 
Similarly large effects are seen on the size measurement. Second, the shape of the error maps for the ellipticity depend on the PSF orientation and  ellipticity, while the error maps for the size show no obvious dependence on the intrinsic PSF ellipticity. 

The above test suggests that the presence of binary stars can have an important impact on the PSF reconstruction. They either must be avoided or their effect must be accounted for in the error budget of the mission. While in principle the Gaia catalogue is expected to be complete up to $V_\text{lim}\sim 20-25$ mag \citep{Robin2012}, binary and multiple stars catalogues will not be complete for the small separations in which we are interested in this work \citep{Bruijne2015}.
In the following, we investigate, both analytically and with image simulations, the effect of unidentified binary stars on the PSF reconstruction for a Euclid-like telescope. 
%
\section{Expected deviations and reconstruction error}\label{sec:th}

\subsection{General case} 

\begin{figure}[h]
 \centering
  \includegraphics[width=0.79\linewidth]{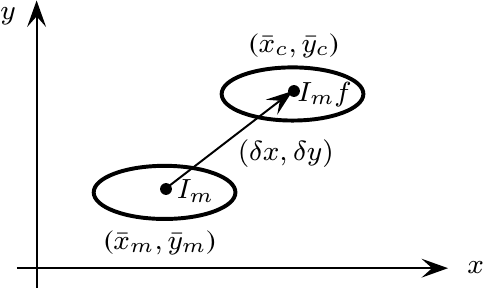}
  \caption{Sketch of the system with the main object at $\bar{x}_m,\bar{y}_m$, with total intensity $I_m$ and the companion located $(\delta x, \delta y)$ away with a total intensity of $I_mf = I_c$.}

\end{figure}

We investigate in an analytical manner how the PSF is altered by the presence of multiple stellar signals.
We show that (i) the bias on the complex ellipticities is a function of both the characteristics of the multiple stars distribution and of the intrinsic complex ellipticities, and that (ii) similarly, the bias on the size is a function of the multiple stars distribution characteristics and of the intrinsic size.

Let $q_{xx}$ be the quadratic moments along the $x$ axis of the image of a star, $q_{yy}$ along the $y$ axis and $q_{xy}$ along the $x-y$ diagonal,
\begin{equation}
  q_{jk}=\frac{\int (j-\bar{j})(k-\bar{k}) \mathrm{I}(x, y)\dx\dy}{\int \mathrm{I}(x, y)\dx\dy},
\end{equation}
where $j, k$ are either $x$ or $y$. The parameter $\bar a$ denotes the position of the centroid for the axis $a$.
The image is centred on the main star, which  is at $(\bar{x}_0,\bar{y}_0)$. 
In a noise-free image, the complex ellipticity components are obtained via \citep{Kaiser1995}:
\begin{equation}
  [e_1, e_2] =\left[\frac{q_{xx} - q_{yy}}{q_{xx} + q_{yy}}, \frac{q_{xy}+q_{yx}}{q_{xx} + q_{yy}}\right] = \left[\frac{q_{xx} - q_{yy}}{q_{xx} + q_{yy}}, \frac{2q_{xy}}{q_{xx} + q_{yy}}\right].
\end{equation}

We compute the expectation values for the quadratic moments to mimic the result of the stacking of many binary systems. We restrict this analysis to the case of the same intensity for all main stars, while each companion may have a different intensity ratio $f_k$. 
All companion stars are randomly distributed around the main star. To get the expectation value of the quadratic moments, we use the limit case when the number of systems in the stack $N_\star \longrightarrow \infty$. We assume that the light profile of the companion star is given by $\mathrm{I}_c(x,y) = f \mathrm{I}_m(x-\delta x,y-\delta y)$ and total intensity of $I_c=fI_m$.

For a binary system, the centroid in the $x$ axis is located at
\begin{eqnarray}
 \bar{x}= \frac{I_m \bar{x}_m + I_c \bar{x}_c}{I_m + I_c}= \bar{x}_m + \frac{f\delta x}{1 + f}.
\end{eqnarray}
The computations are analogous for the $y$ axis. We compute the total quadratic moment of the system $k$,  $q_{xx,k}$, in the $xx$ direction,  via the parallel axis theorem. The quadratic moment of the binary system $k$ along $xx$ is denoted $q_{xx,m}$, i.e.
\begin{eqnarray}\label{eq:qxxtot}
q_{xx,k} &=& \frac{q_{xx,m} + (\bar{x}_m - \bar{x})^2
              + f_kq_{xx,m} + f_k(\bar{x}_c - \bar{x})^2}{1+f_k}
              \nonumber\\
              & = & q_{xx,m} + \frac{f_k^2 \delta x^2}{(1+f_k)^3} + \frac{f_k \delta x^2}{(1+f_k)^3}
              \nonumber\\
              & = & q_{xx,m} + \frac{f_k\delta x_k^2}{(1+f_k)^2}.
\end{eqnarray}
The derivation for $q_{yy,k}$ is similar. 
\subsubsection{Biases for complex ellipticities}
The position of the companion stars are random, which implies that $\langle \delta x^2 \rangle = \langle \delta y^2 \rangle$, where the notation $\langle \cdot \rangle$ denotes the expectation value. The expectation value for the complex ellipticity component $e_1$ is
\begin{equation} \label{eq:e1tot}
 \langle e_{1} \rangle =\frac{
q_{xx}-q_{yy}}{
q_{xx}+q_{yy} + \langle f r^2/(1+f)^2 \rangle}.
 \end{equation}
As the term $\langle f r^2/(1+f)^2 \rangle$ is close to zero in Eq. \ref{eq:e1tot}, its Taylor expansion is taken and we get the value of the bias for the system~$k,$
\begin{equation}
 \delta e_{1,k} =  e_{1,k}-e_{i,0},
\end{equation}
for $i=1,2$ and $e_{i,0}$ the $i$th component of the complex ellipticity, such that the expectation value for the bias is
\begin{equation} \label{eq:e1}
  \langle \delta e_{1}\rangle \simeq -e_{1,0}\frac{1}{R^2_0} \left\langle\frac{ r^2 f}{(1+f)^2}\right\rangle,
\end{equation}
where $R_0^2$ is the intrinsic size of the PSF defined as the sum of the quadratic moments in the $xx$ direction plus the moments in the $yy$ direction. This quantity is referred to as $T$ in \citet{Kaiser1995}. Similarly, for $e_{2}$,

\begin{equation} \label{eq:e2}
 \langle \delta e_{2}\rangle  \simeq -e_{2,0} \frac{1}{R^2_0}  \left\langle\frac{ r^2 f}{(1+f)^2}\right\rangle.
\end{equation}
The above set of equations shows that the bias, $\delta e_{i}$, is a function of the intrinsic ellipticity of the PSF, $e_{i,0}$, and of the companion distribution through the quantity $\langle r^2 f / (1 + f)^2 \rangle$. It also shows that the magnitude of the bias is the same both in $e_1$ or $e_2$ components. 
The intrinsic size of the PSF $R_0^2$ can mitigate the effect of the binaries. If we use the same binary population  for small and large PSFs (i.e. ensuring that the same binaries are excluded in both cases), the bias is smaller than for small PSFs.
Another conclusion drawn from the presence of a minus sign is that the presence of binaries in the stack tends to make the reconstructed PSF rounder than the intrinsic PSF shape.
\begin{table*}
\caption{\label{tab:DK13}Multiplicity properties for population I main-sequence stars used for this study (adapted from \citetalias[][]{Duchene2013}).}
\centering
\begin{tabular}{cccc}
\hline\hline 
Mass range & Mult./comp. & Mass ratio & Orbital period\\
 & frequency & distribution & distribution \\
\hline
\multirow{2}{*}{$M_\star \lesssim 0.1\,M_\odot$} & $MF = 22^{+6}_{-4}\,\%$ & \multirow{2}{*}{$\gamma = 4.2 \pm 1.0$} & Unimodal (log-normal?) \\
 & $CF = 22^{+6}_{-4}\,\%$ & & $\overline{a} \approx 4.5$\,AU, $\sigma_{\log P} \approx 0.5$ \\
\hline
\multirow{2}{*}{$0.1\,M_\odot \lesssim M_\star \lesssim 0.5\,M_\odot$} & $MF = 26\pm3\,\%$ & \multirow{2}{*}{$\gamma = 0.4 \pm 0.2$} & Unimodal (log-normal?) \\
 & $CF = 33\pm5\,\%$ & & $\overline{a} \approx 5.3$\,AU, $\sigma_{\log P} \approx 1.3$ \\
\hline
\multirow{2}{*}{$0.7\,M_\odot \lesssim M_\star \lesssim 1.3\,M_\odot$} & $MF = 44\pm2\,\%$ & \multirow{2}{*}{$\gamma = 0.3 \pm 0.1$} & Unimodal (log-normal) \\
 & $CF = 62\pm3\,\%$ & & $\overline{a} \approx 45$\,AU, $\sigma_{\log P} \approx 2.3$ \\
\hline
\multirow{2}{*}{$1.5\,M_\odot \lesssim M_\star \lesssim 5\,M_\odot$} & $MF \geq 50\%$ & \multirow{2}{*}{$\gamma = -0.5 \pm 0.2$} & Bimodal \\
 & $CF = 100\pm10\,\%$ & & $\overline{P} \approx 10$d \& $\overline{a} \approx 350$\,AU \\
\hline
\end{tabular}
\tablefoot{Multiple systems are characterised by the frequency of multiple system $MF$ and the companion frequency $CF$, i.e. the average number of companions per star. The latter can exceed 1. The two last columns present the parameters for  mass ratio distribution and for  orbital period distribution. See \citetalias[][]{Duchene2013} for more details.}
\end{table*}
\subsubsection{Bias for size}
The size of the PSF is $R^2 = q_{xx} + q_{yy}$. The bias on the size is written $\delta R^2/R_0^2 = \left( R^2  - R^2_0\right)/R_0^2$. 
Computing the bias gives
\begin{eqnarray} \label{eq:deltar}
 \frac{\delta R^2}{R_0^2} &\simeq& \frac{1}{R^2_0}\left\langle\frac{  r^2 f}{ (1+f)^2}\right\rangle.
\end{eqnarray}
Thus changes in size depend on the binary star fraction distribution and not on the intrinsic PSF ellipticity.
%

\subsection{Ideal reconstruction error}
The reconstruction error can be bounded by a requirement for future weak lensing surveys. We write the reconstruction error as
\begin{equation} \label{eq:sigmae}
 \sigma(e_i) = \sqrt{\frac{1}{N_s-1}\sum_{s=1}^{{N_s}} \left(e_{i,s} - e_{i,0}\right)^2},
\end{equation}
where $s=1\dots N_s$ runs over the $N_s$ stacked images of stars and $e_{i,s},e_{i,0}$ are the ellipticity components of the resulting stacked image and of the ideally reconstructed PSF, respectively. 
If $N_s$ is large enough, we can write
\begin{equation}\label{eq:sigma_e}
 \sigma(e_i) =  \langle e_{1}\rangle - e_{i,0} \simeq \frac{1}{R^2_0}\left\langle\frac{  r^2 f}{ (1+f)^2}\right\rangle e_{i,0}.
\end{equation}
Similarly, for the size
\begin{equation} \label{eq:sigmar}
 \sigma(\delta R^2)/R_0^2 \simeq \frac{1}{R^2_0}\left\langle\frac{  r^2 f}{ (1+f)^2}\right\rangle.
\end{equation}

\subsection{Suboptimal reconstruction error} 
\label{sec:suberror}

The above equations are derived for an image that contains a very large number of co-added stellar systems $N_\star$. If this number gets too small, shot-noise error dominates, introducing an extra source dispersion in the reconstruction error $\sigma$. 
When $N_\star$ is too low, the expectation value $\langle r^2f/(1+f)^2\rangle$ will no longer be representative of the stellar binary distribution in the image. We discuss the minimal value of $N_\star$ to escape this suboptimal regime in Sect.~\ref{nstar}.

The separation $r$ and ratio of the intensity $f$ distributions are typically log-normal as described in \citetalias[][]{Duchene2013}. 
The bias $\delta e$ is the multiplication of those two log-normal distributions, thus the bias is also an exponential-like distribution. 
The standard deviation of such exponential-like distributions has a $\sigma_N \sim 1/\sqrt{N_\star}$ dependence \citep[e.g.][]{Hoogenboom2006}.  At low $N_\star$, this $\sigma_N$ is more important than the reconstruction error $\sigma(x),$ where $x$ is either the ellipticity component $e_i$ or the size $R^2$. The parameter $\sigma(x)$ is the binary PSF reconstruction error signal, while $\sigma_N$ is a statistical artefact. 
The total measured error $\sigma_\text{meas}$ is a combination of $\sigma_N$ and $\sigma(x)$.
For low $N_\star$ or low binary PSF reconstruction errors, the statistical error $\sigma_N$ dominates the reconstruction error.
On the other hand, when enough binaries are stacked or when the PSF properties exhibit a large value of the size or the ellipticity, the binary PSF reconstruction error $\sigma(x)$ dominates.
%
\section{Binary stars in the Milky Way} \label{sec:binaries}
%
\begin{figure*}
 \centering
  \includegraphics[width=0.48\linewidth]{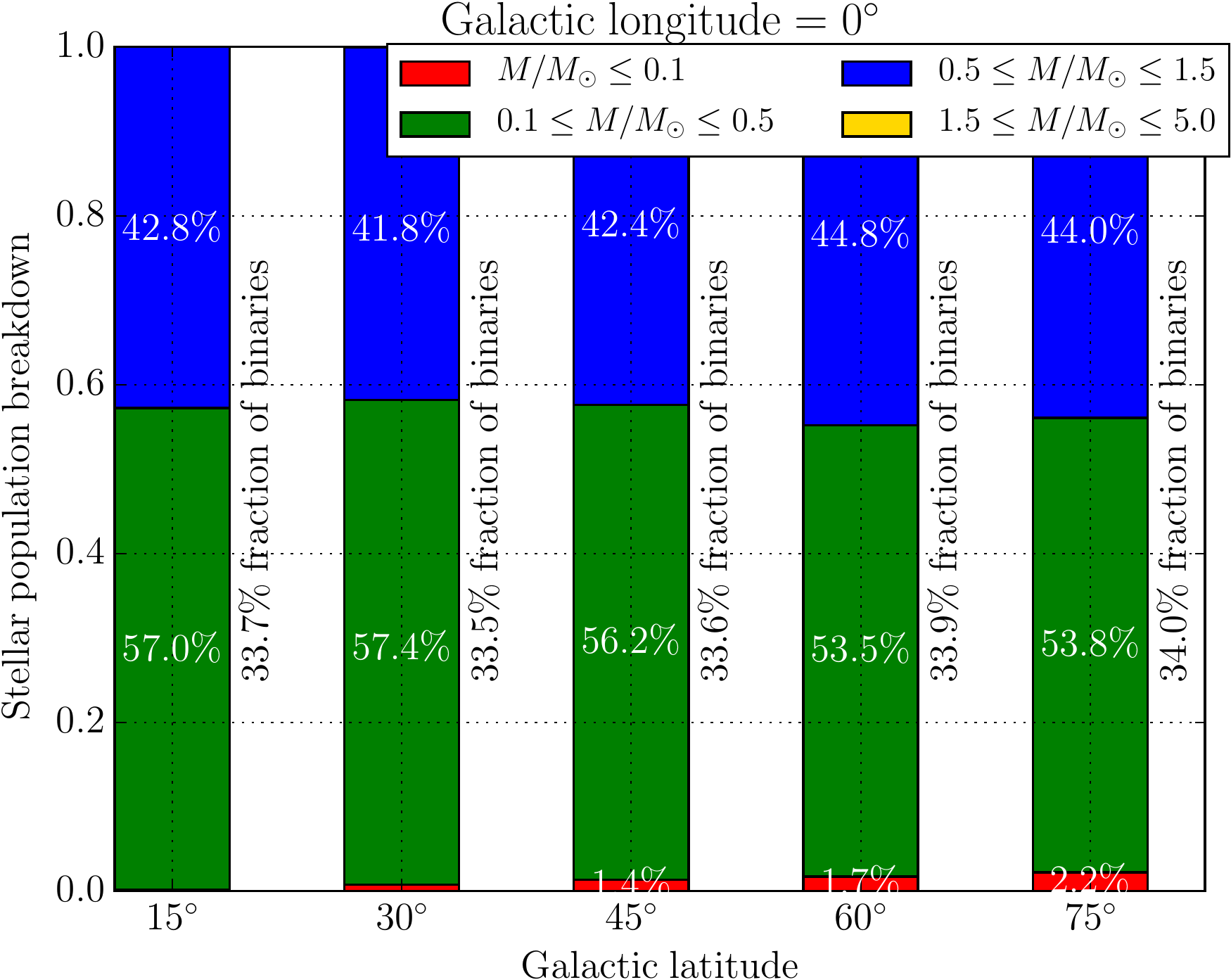}\hskip 15pt
  \includegraphics[width=0.48\linewidth]{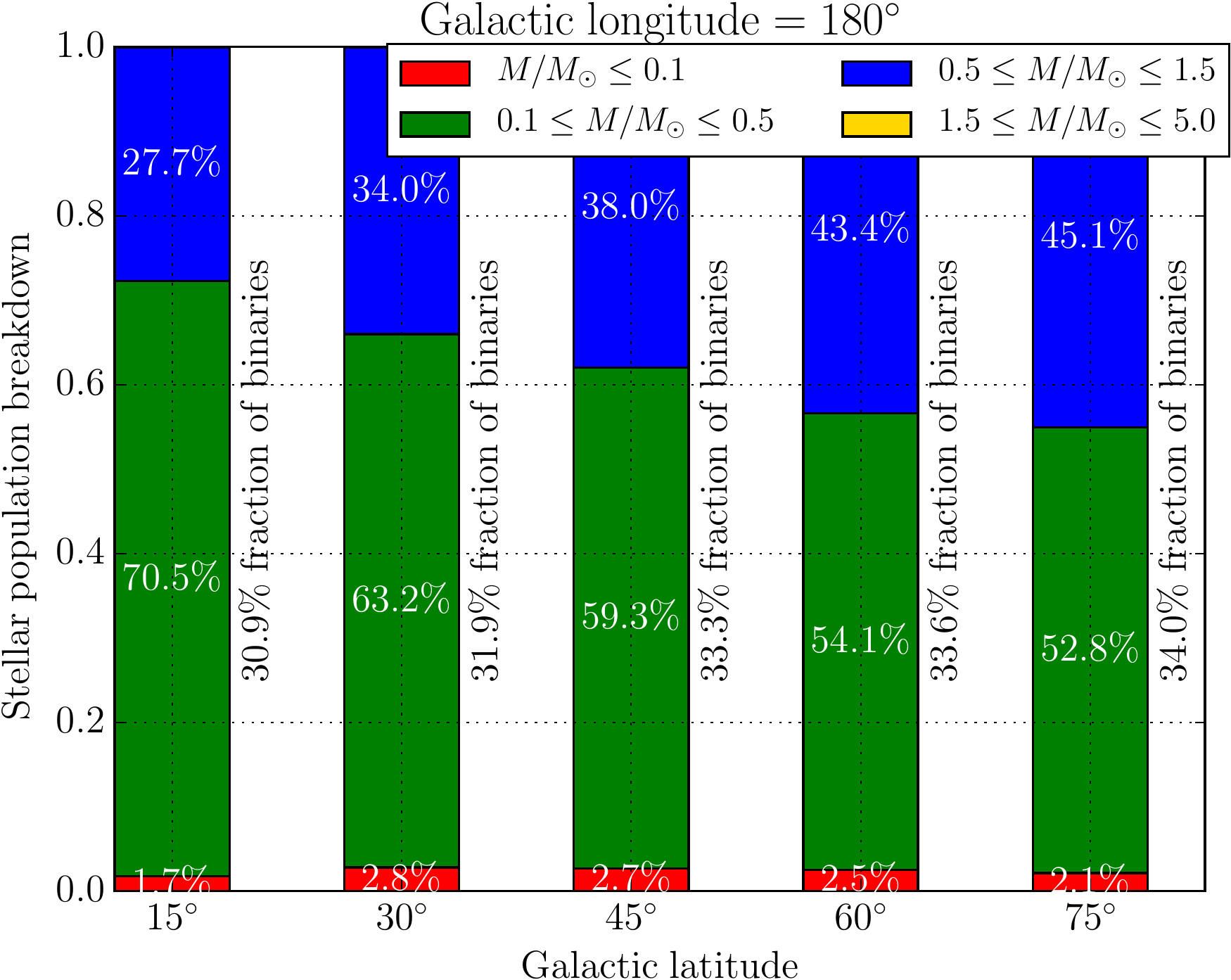}
  \caption{\label{fig:popana}Main-sequence stellar population breakdown for different Galactic latitudes. The Galactic longitude on the left panel is $\ell=0^\circ$ (towards the centre of the disk) and the right panel $\ell=180^\circ$ (Galactic anti-centre). The bin of massive stars ($>1.5~M_\odot$) is not visible as it represent less than 0.2\%. We stress the different behaviour of the binary fraction for massive stars (blue vs green bins) for the opposite directions in Galactic longitude.
  }
\end{figure*}

We now test our predictions for the PSF reconstruction errors. To accomplish this, we draw a realistic population of stars from the Besan\c{c}on Model of the Galaxy \citep[BMG;][]{Robin2004}. To this population of single stars, we add companions according to the properties of binary stars described in \cite[][hereafter DK13]{Duchene2013}, including the orbital and mass ratio parameters. We simu\-late the PSF reconstruction process by stacking many observations of a mix of stellar systems with and without companions. We then measure the ellipticity and size of the stacked light profile and compare them with the intrinsic values in the absence of any stellar companion.

In the following section, we present the properties of the multiple systems in the Milky Way and   discuss the demographics of binary stars with respect to their position on the sky before we describe the simulations themselves.

\subsection{Characteristics of multiple stellar systems}
\label{sec:CharacteristicsMS}

\citet[][]{Duchene2013} found that multiple systems are characterised by two main quantities: (i) the frequency of multiple system, $MF$, i.e. the number of occurrences of a binary system in a given population; and (ii) the companion frequency, $CF$, the average number of companions per star, which can exceed one. 
We restrict ourself to binary stars and only use the frequency of multiple systems, $MF$, to draw populations of binary stars.
The physical separation distributions follow a log-normal curve. The detailed representation of the log-normal distribution function suggests different formation processes for multiple systems \citep[e.g.][]{Heacox1996}.
The mass ratio of the companion mass to the primary mass, $q\leq1$, can be determined from the observation of the flux ratio assuming mass-luminosity relations. The distributions of the mass ratio follow an empirical power law. Table~\ref{tab:DK13} summarises important trends in the population of binary systems. Some of the main trends are:
\begin{enumerate}
 \item The more massive the primary star, the larger the probability of a companion,
 \item The more massive the primary star, the lower the mass ratio power-law index $\gamma$. This implies that massive stars tend to host companions of lower relative masses than low-mass stars,
 \item The more massive the primary star, the broader the physical separation with the companion.
\end{enumerate}
%
\subsection{Spatial distribution of binaries}

The demographics of stellar populations depend on the Galactic coordinates and most notably on the Galactic latitude, and since different regions of the Milky Way have different stellar populations, they have different fractions of multiple systems. This fraction is mainly controlled by the fraction of massive stars. An informed choice of the stellar population used for the PSF reconstruction can improve the purity of the sample by minimising the fraction of multiple systems.

We illustrate this effect in the following with several realisations of the BMG main-sequence stars at different Galactic coordinates. For each simulation, the number of binary stars is evaluated. 
Figure~\ref{fig:popana} shows two different examples for Galactic longitude $\ell=0^\circ$, near the Galactic centre and $\ell=180^\circ$, in the direction of the Galactic anti-centre. 

Towards the Galactic centre, the stellar population demographics remain approximately constant for latitudes between $b=15^\circ$ and $b=75^\circ$. Thus, there is no significant variation in the fraction of binary stars. On the other hand, pointing close to the anti-centre results in an increasing fraction of massive stars with increasing Galactic latitude. 
Stars of masses between 0.5~$M_\odot$ and 1.5~$M_\odot$ represent less than $30\%$ of the stellar population at $b=15^\circ$, while they account for almost half of the population at $b=75^\circ$. 
When this observation is combined with the binary star information of Table~\ref{tab:DK13}, this leads to a statistically significant increase of the fraction of binary stars with increasing Galactic latitude, as shown in Fig.~\ref{fig:popana}. This means that the impact of the multiplicity of stars on the PSF determination is not the same at different locations on the sky, not only because of the projected number density of stars but also because of the varying stellar populations. It also means that PSF calibration fields must be carefully chosen and the fraction of multiple stars in the selected fields must be accounted for.
In the following, we choose the pointing $\ell=180^\circ, b=45^\circ$, as the population demographics and fraction of binary stars seems to be representative of all available data and because this pointing is located at a portion of the sky surveyed by Euclid \citep{Euclid}.

\subsection{Synthetic images of single and binary stars} \label{sec:ssp}

Our goal is to measure the PSF reconstruction errors for a given number of stars that can be co-added but that are sometimes binaries. To this end, several simulations with different fractions of binary stars are generated. 
The stars potentially useful for the PSF reconstruction of future space-based surveys have magnitudes in the range $i(AB)\sim 16-24$, depending on the exposure duration and detector saturation levels. 
Thus, we only keep  stars falling into this range  in the sample. Main-sequence stars more massive than $1.5$ solar masses are discarded as well owing to the lack of data about their multiplicity. However, these stars represent less than 0.2\%\ of the whole dataset. We also discard stars that have more than one companion. All our simulations are therefore on the optimistic side in this respect. 

The stellar populations generated by the BMG do not contain stellar companion information. We therefore add companions to the BMG stars using the information in \citetalias[][]{Duchene2013} and summarised in Table~\ref{tab:DK13}. We replace missing values with the value in the lowest bin of mass, which represents a conservative estimate.

A mass-luminosity relation for low-mass main-sequence stars is used to turn the mass of the companion into a magnitude \citep{Habets1981}. For simplicity, we assume that all the stars have the same flat spectral energy distribution. In other words, our simulated images are for flat SEDs but the allocation of companions to the stars follows the prescription of \citetalias[][]{Duchene2013} applied to the stellar populations of the BMG. 

The position of a companion star along the assumed circular orbit is drawn from a uniform distribution. The angle between the plane of the orbit and the line of sight, as well as the longitude of the ascending nodes, are also uniformly distributed.
The light profile of the companion is normalised according to the mass-luminosity relation, and interpolated and shifted by the required vector. 
Magnitudes of both objects are modified such that the total apparent magnitude of the system is the original apparent magnitude given in the BMG catalogue.
We account for the large uncertainties in the stellar multiplicity parameters and explore the behaviour of the PSF reconstruction at different binary star fraction,  by randomly modifying the total binary star fraction of the stellar population  between each realisation of the catalogue to yield a fraction of binary stars between 0 and 1.

When building images of stars and  their companions, we adopt a spatial sampling 12 times better than the expected survey resolution. We simulate Euclid-like images that have the same resolution as the VIS camera, using Gaussian PSFs with a FWHM of 0.137\arcsec. In doing so, we adopt a pixel size of 0.0083\arcsec\ and a filter bandpass that matches the VIS instrument of Euclid. Down-sampling of the images to the same pixel size as the actual data is not necessary and has little impact on the results in the absence of noise. The results we  present in the rest of the paper do not depend on the details of the shape of the PSF, i.e. Gaussians give the same results as more realistic diffraction-limited PSFs generated with the {\tt GalSim} software \citep{Rowe2015}. This is mainly because the effect of unresolved stellar companions affects the very centre of the PSF.

When dealing with real (noisy) data, the PSF is best reco\-vered by simultaneously considering many different point sources to improve the signal to noise. To mimic this process, we stack many 
images of stars (some containing binaries) generated as described above. In the present case, all images are (i) noise-free and (ii) centred on the measured centroid of the binary system or the centroid of the single star. 
Stars are only stacked within a narrow magnitude range. Binaries separated by 0.05\arcsec are removed from the sample, i.e. we assume that in a real survey these binaries can be identified as such and removed from the stars used to build the PSF.
Each image of a stack is composed of a mix of single and binary stars with realistic separations and contrasts.
Measurements of the ellipticity and the size of the stacks are computed with adaptive moments by \citet{Hirata2003}, as implemented in {\tt GalSim} \citep{Rowe2015}.

A large number of stacks are measured and divided into bins of similar binary fractions.
The biases and random errors on the reconstructed size and complex ellipticities are then estimated by comparing with the input values in the absence of binaries. 
For each group of stars with a similar binary fraction we measure the standard deviation for the sizes and ellipticities of all objects in the group. This gives an estimate of the precision achieved on a given parameter. The bias is the standard deviation of the difference between the ellipticity of the stack and the input values in the absence of binaries. In the rest of the paper we use the following terminology:

\begin{itemize}
 \item A stack of stars is an ensemble of stars that all have the same PSF shape and are drawn from the BMG using the binary properties of \citetalias[][]{Duchene2013}.
 \item $N_\star$ is the number of stars in each stack; a binary system is counted only once.
 \item $N_s$ is the number of realisations of the stacks generated for each simulation. Each stack contains $N_\star$ stars drawn $N_s$ times. $N_s$ is always large enough to make the numerical errors on the simulation negligible in front the parameters that are reconstructed, i.e. the size and ellipticity of the PSF. Typical values for $N_s$ are hundreds of thousands of realisations.
 \item PSF reconstruction refers to the measurement of the PSF profile of the stack (size and ellipticity).
 \item The error on the reconstructed parameters are defined in Eqs.~\ref{eq:sigmae} and \ref{eq:sigmar}.
 \item The bias on the reconstructed parameters are defined in Eqs.~\ref{eq:e1}, \ref{eq:e2}, and \ref{eq:deltar}.
\end{itemize}
%
%
\section{Results} 
\label{sec:discussion}
\begin{figure*}
 \centering
  \includegraphics[width=.49\linewidth]{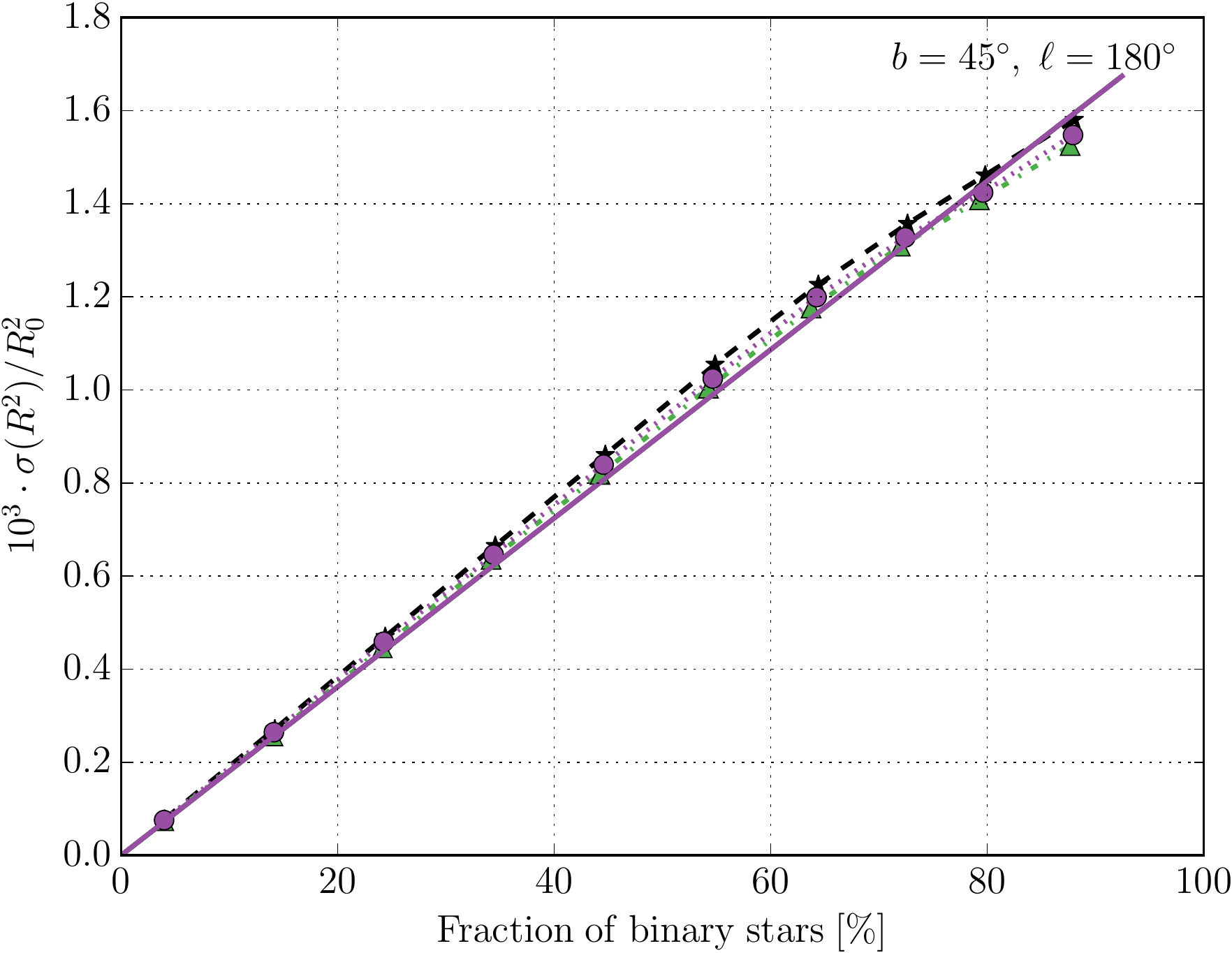}
  \includegraphics[width=.49\linewidth]{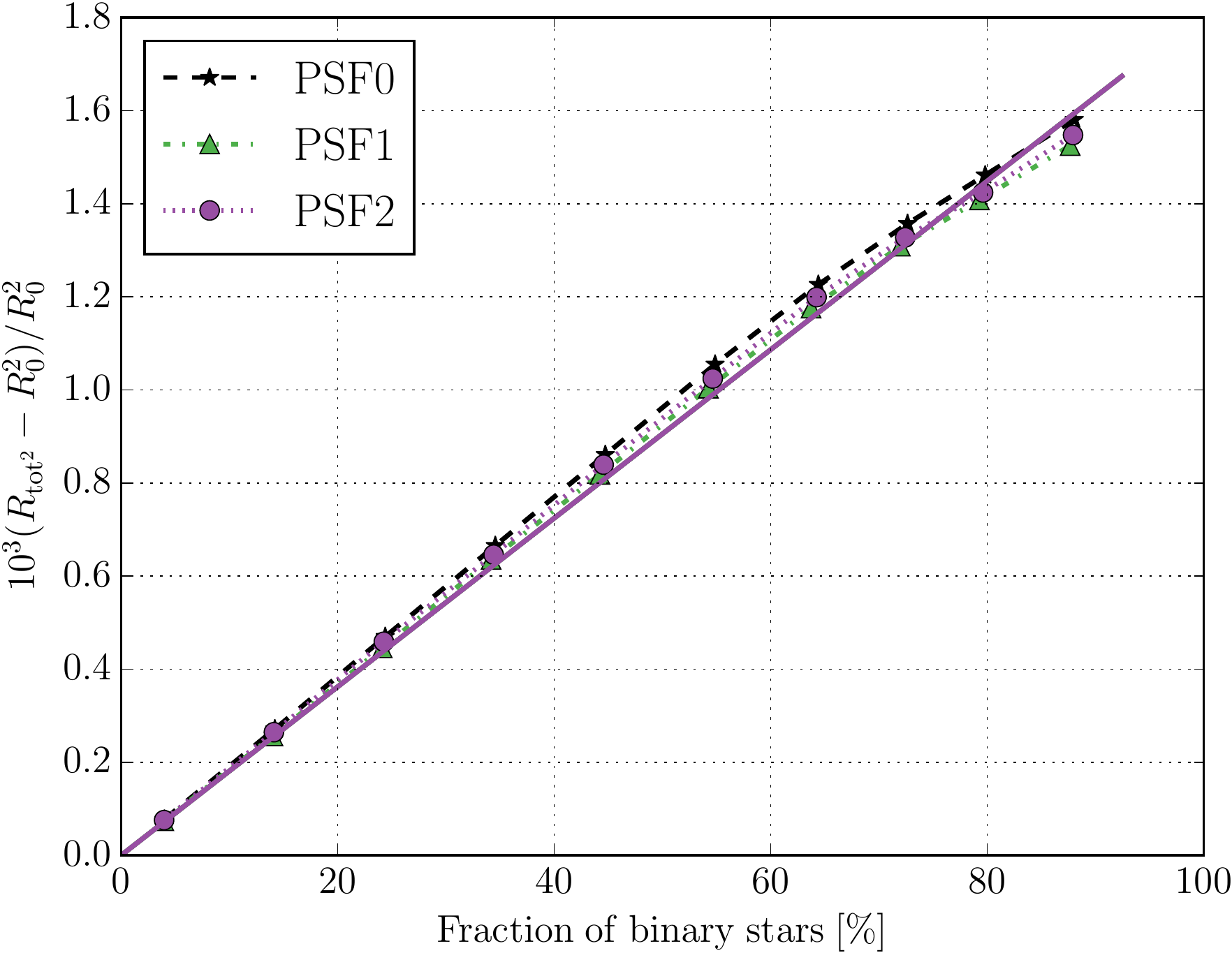}
  \vskip 20pt
    \includegraphics[width=.49\linewidth]{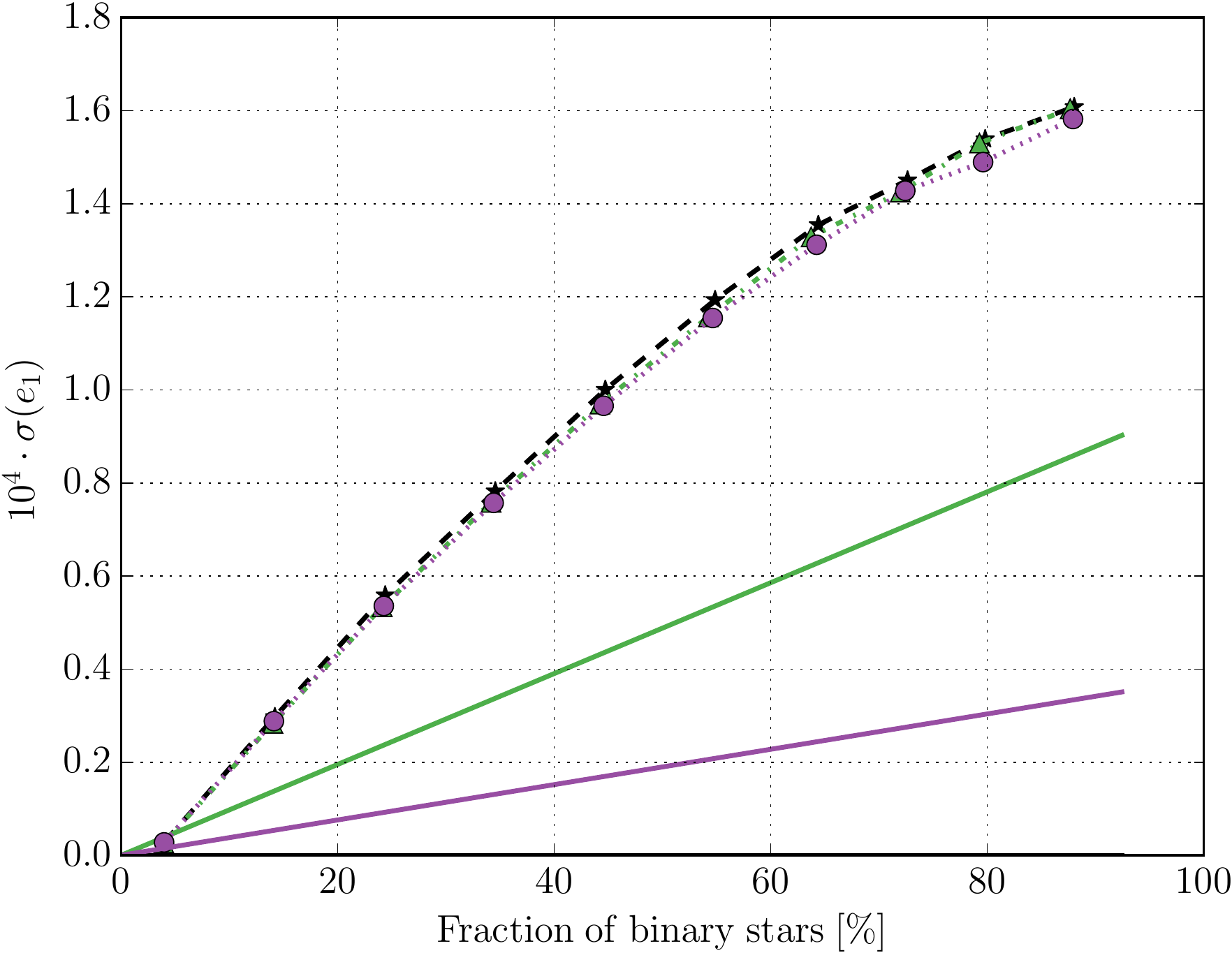}
  \includegraphics[width=.49\linewidth]{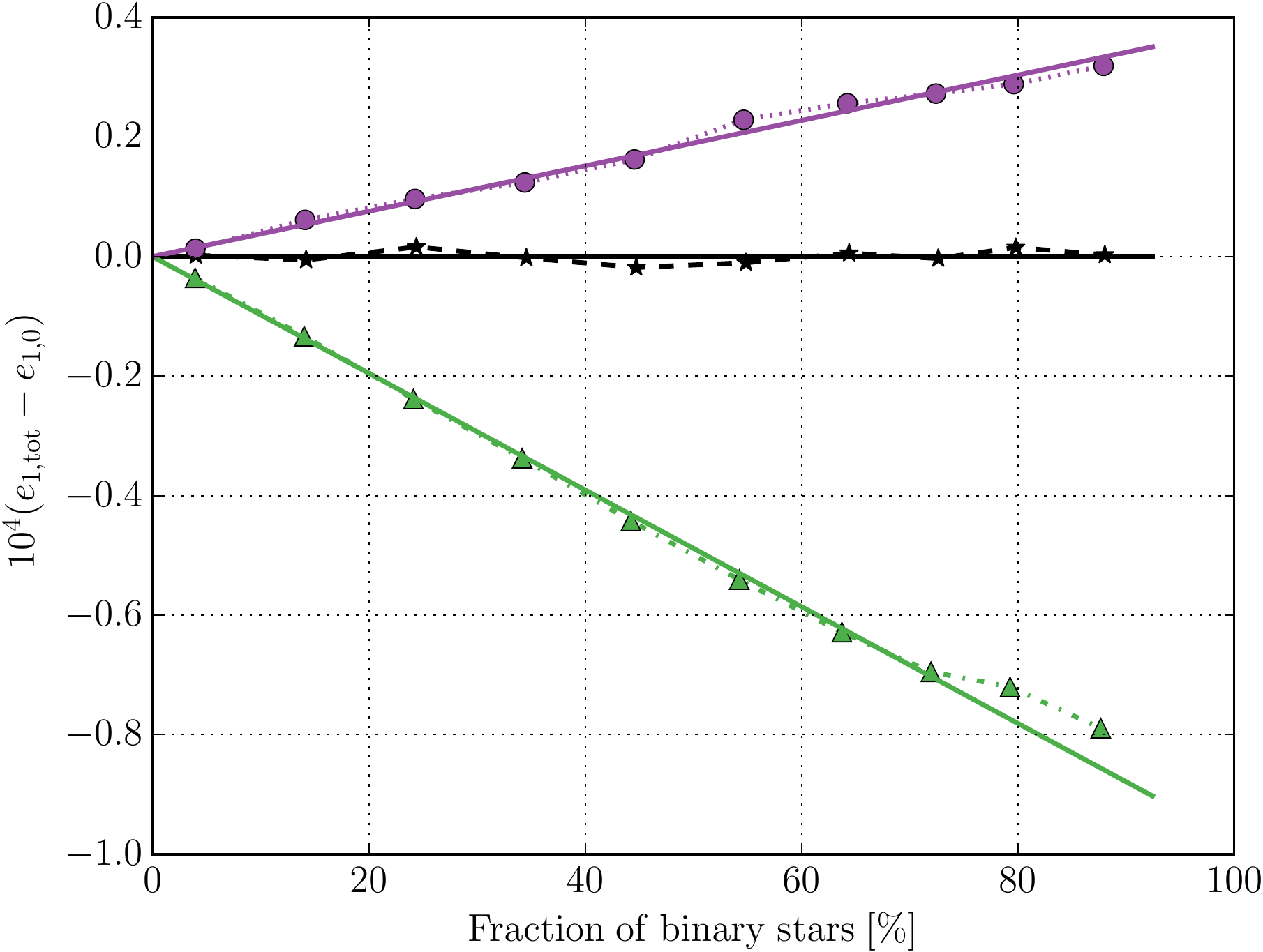}
  \vskip 20pt
     \includegraphics[width=.49\linewidth]{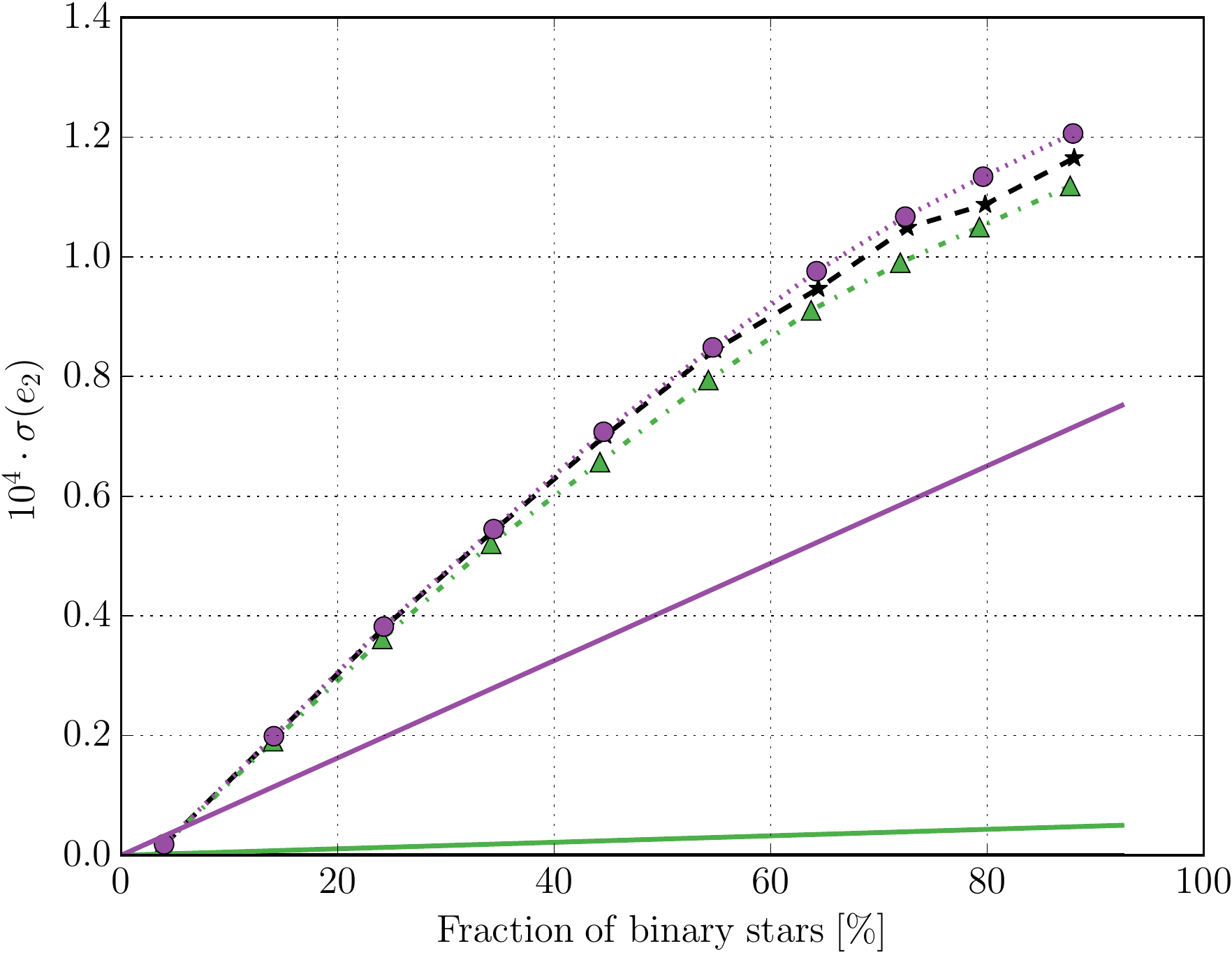}
   \includegraphics[width=.49\linewidth]{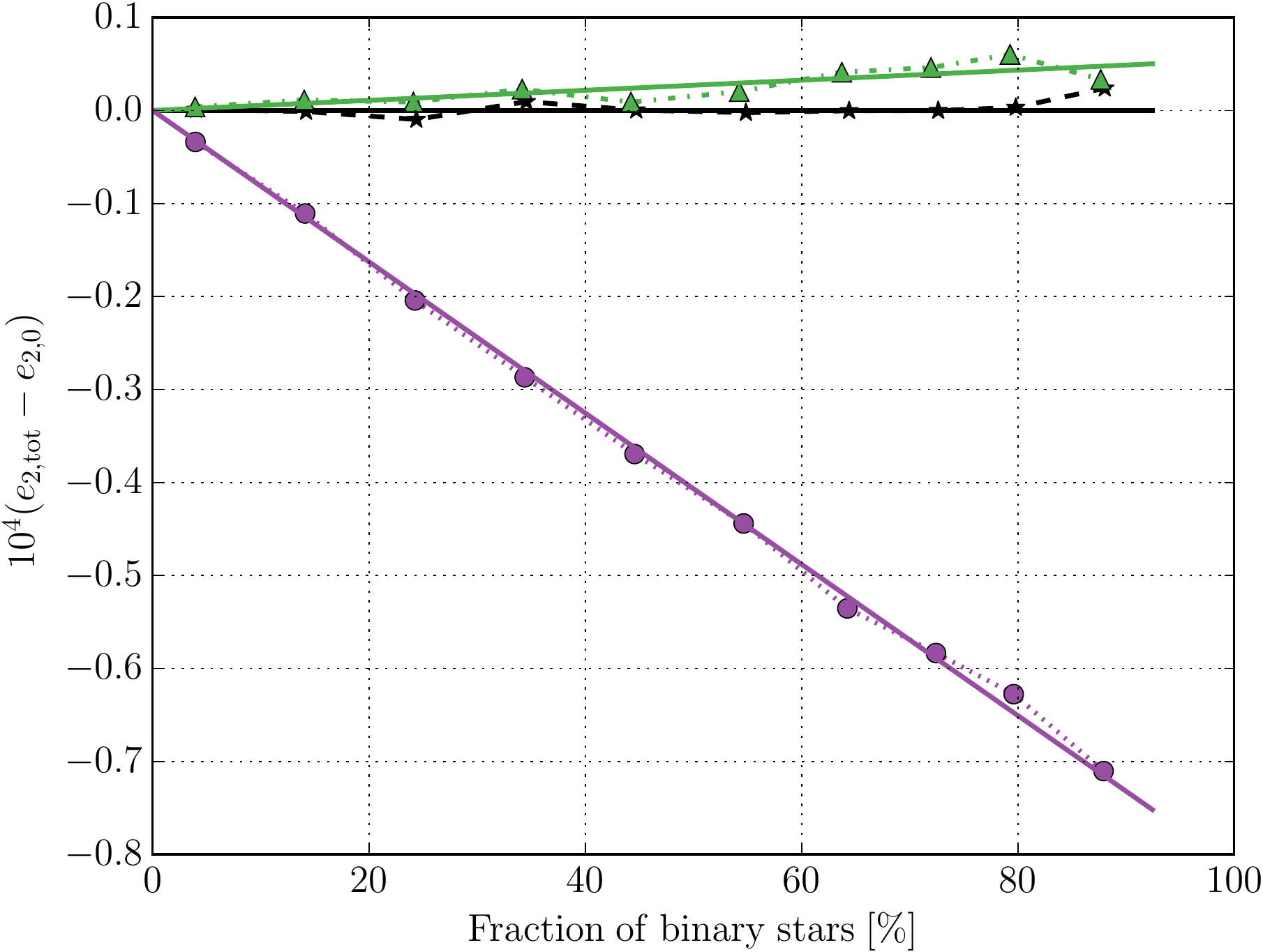}
  \caption{PSF reconstruction errors (\emph{left column}) and biases (\emph{right column}). From top to bottom,  plots for the size and  two ellipticity components, $e_1$ and $e_2$, are shown. In each panel, the values are given as a function of the binary fraction in the stellar population considered and for the three different PSF shapes described in Table \ref{tab:PSF}. The solid lines represent the predicted values using the formalism in Sect.~\ref{sec:th}. The latter is valid for $N_\star \rightarrow \infty$ while the simulations used in this figure are for $N_\star=40$ stars with apparent magnitude in the range $18\leq i(AB) \leq 19$.}
  \label{fig:gauss_errstd}
\end{figure*}
In the following, we analyse the simulations described in Sect.~\ref{sec:ssp}. We focus on how the error in the PSF reconstruction due to multiple star blending varies with the intrinsic PSF ellipticity, i.e. in the absence of blends, and we show our results as a function of the total fraction of binary stars. We carry out the experiment both for Gaussian PSFs and for diffraction limited PSFs. 
We then explore the impact of the number of stars, $N_\star$, available to reconstruct the PSF. Finally, we show the effect of changing the stellar populations, i.e. the magnitude distribution of the stars contained in each stack.

%
\subsection{Varying the fraction of binaries for different PSF}
\label{binfrac}

\begin{figure*}
 \centering
  \includegraphics[width=.49\linewidth]{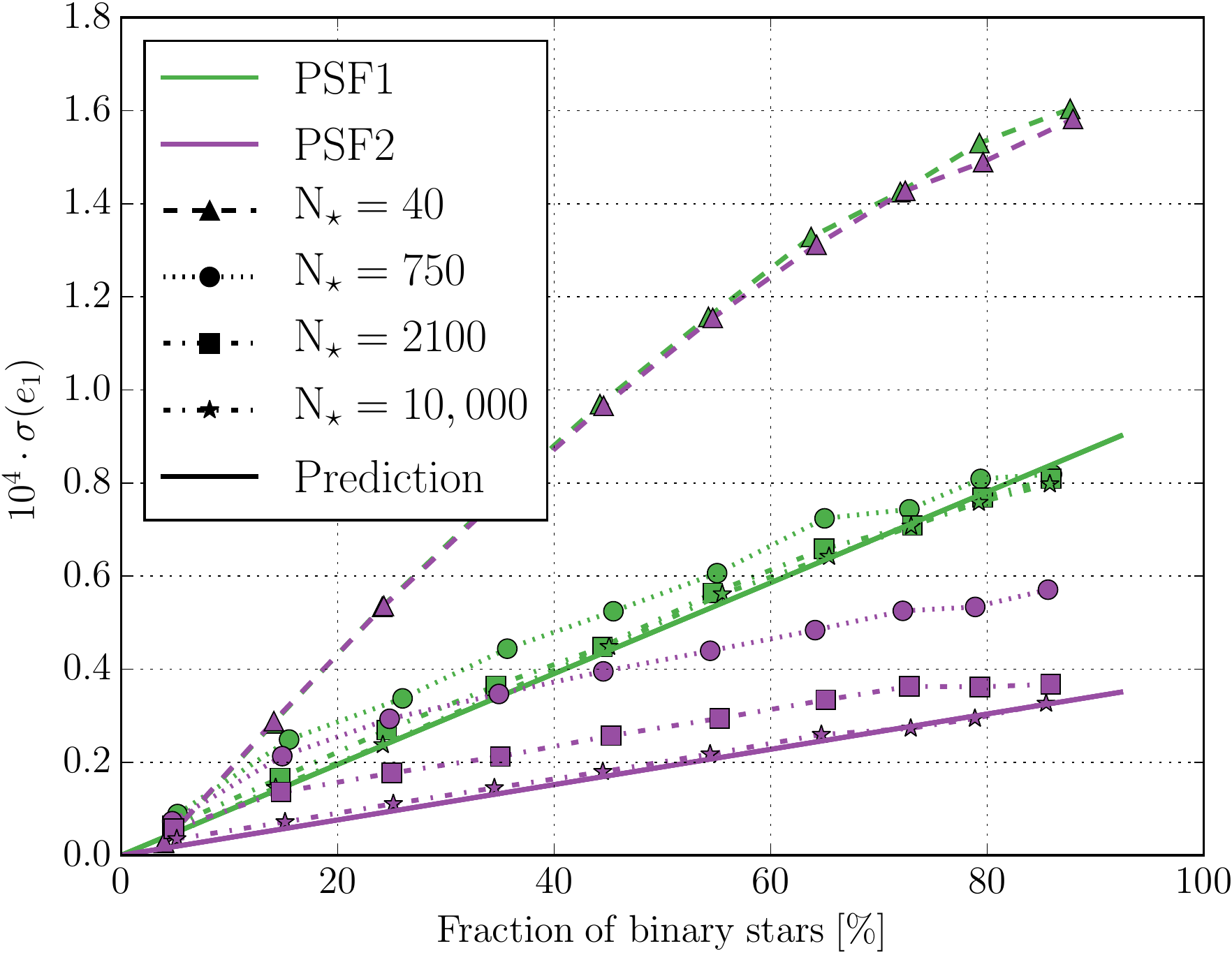}
  \includegraphics[width=.49\linewidth]{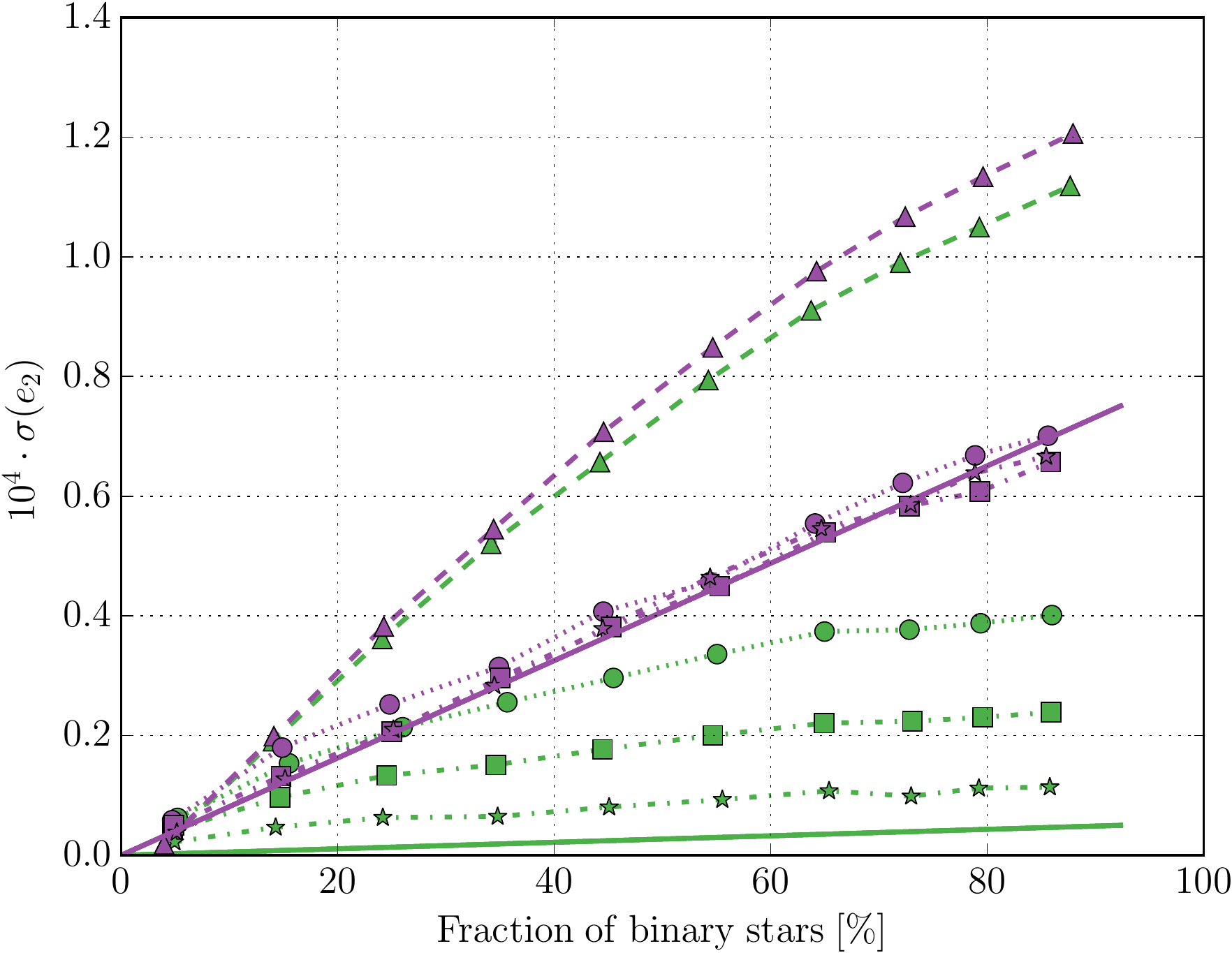} 
  \caption{PSF reconstruction errors on the two ellipticity components, $e_1$ (\emph{left}) and $e_2$ (\emph{right}). These plots correspond to the two panels in the lower left corner of Fig.~\ref{fig:gauss_errstd}, but we now show the effect of changing the number of stars, $N_\star$, used in the stack. Depending on the intrinsic ellipticity of the PSF (see Table~\ref{tab:PSF}), thousands of stars are needed to avoid being dominated by the shot noise. \label{fig:proba_std}}
\end{figure*}

We design a simple numerical experiment using three different Gaussian PSFs with the properties shown in Table~\ref{tab:PSF}. For each PSF, we generate ensembles of stars with different fractions of binaries and we produce noise-free stamp images on which we measure the ellipticity and  size using a simple moment calculation. 

Our results are summarised in Fig.~\ref{fig:gauss_errstd}, where we choose $N_\star=40$ as an example. In practice, $N_\star$ can change with different survey characteristics, in particular, the temporal and spatial stability of the PSF. In other words, $N_\star$ reflects the number of stars that can be used simultaneously to reconstruct the PSF.

Figure~\ref{fig:gauss_errstd} indicates the reconstruction error for the size, $\sigma(R^2)/R_0^2$ and for the two ellipticity components, $\sigma(e_1)$ and $\sigma(e_2),$ as a function of the fraction of binary stars. We also show the biases on the size and on the ellipticity. In each panel, the figure compares  the errors and  biases, as derived from our simulations (symbols), to the predictions (solid lines) derived in Sect.~\ref{sec:th}. 

\begin{table}[t!]
\caption{\label{tab:PSF} Shape of the three PSFs used in the simulations.}
\centering
\begin{tabular}{lccccl}
\hline\hline 
  ID & $e_1$ & $e_2$ & $e$ & FHWM (\arcsec)& $PA$($^{\circ}$)\\
\hline
PSF0  & +0.000   & +0.000   & 0.000  & 0.137 & +0.0 \\
PSF1  & +0.054 & $-$0.003 & 0.054 & 0.137 & +88.4\\
PSF2  & $-$0.021 & +0.045 & 0.050 & 0.137 & +145.5\\
\hline
\end{tabular}
\tablefoot{For each PSF, we give the ellipticity components and module, FHWM, and  position angle.}
\end{table}

The two upper panels of Fig.~\ref{fig:gauss_errstd} show that the error and  bias on the size as a function of the binary star fraction is independent of the PSF ellipticity, as suggested by Eq.~(\ref{eq:deltar}). As one would expect, increasing the fraction of binary stars in the stacks increases the size of the PSF. 
The increase with the binary fraction is linear as the characteristics of the binary stars are the same in each bin. The only difference between the bins is the probability of observing a binary system. The figure also shows excellent agreement between the theoretical prediction and the simulation, even for the small number of stars in the stacks, here $N_\star=40$.

The reconstruction errors $\sigma(e_1)$ and $\sigma(e_2)$  on the complex ellipticity, as obtained from the simulations, appear very similar. They seemingly do not depend on the value of the intrinsic ellipticity as Eq.~(\ref{eq:sigma_e}) suggests. However, with $N_\star=40$ stars, we fall in the suboptimal reconstruction scheme described in Sect.~\ref{sec:suberror}. 
Those cases are shot-noise-limited because of the small number of stars in the stack, i.e. the PSF reconstruction error of the ellipticity is dominated by  noise due to the small number of random positions for the companions. 
We  show in Sect.~\ref{nstar} that, as is intuitively expected, increasing $N_\star$ reconciles the errors found from the simulations and  theoretical prediction: $\sigma(e_1)$ and $\sigma(e_2)$ do depend on the ellipticity of the PSF following Eq.~(\ref{eq:sigma_e}) in the limit of large values for $N_\star$. 

We now turn to the biases due to stellar binarity. The bias on the size follows the same curve as a function of stellar binarity as the error on the size. The equations describing the bias and the error, as reported in Sect.~\ref{sec:th}, are 
\begin{eqnarray}
 \frac{\delta R^2}{R_0^2} \simeq \frac{\sigma(\delta R^2)}{R_0^2},
\end{eqnarray}
which is also well illustrated with our Gaussian simulations in the upper right panel of Fig.~\ref{fig:gauss_errstd}.  The bias, both on the size and ellipticity, is easily measured from simulations with $N_\star=40$ in contrast to the measurement of the errors, which require much larger values for $N_\star$. 

The behaviour of the ellipticity is not the same as for the size. For a round PSF with $e_1 = e_2 = 0$, both the bias and reconstruction error equals zero in the limit where $N_\star \rightarrow \infty$, as the net effect of adding round PSFs at random angular positions around a star is null.
The remaining bias for a round PSF, both on $e_1$ and $e_2$, is of the order of $2 \times 10^{-6}$. This can be attributed to the effect of a large but finite number of stacks $N_s$ and to numerical errors. 

For non-circular PSFs, the bias on $e_1$ and $e_2$ shows a linear correlation to the intrinsic complex ellipticity of the PSF, as does the error. The bias and  reconstructions on the ellipticity have the same analytical form with opposite signs, and both are proportional to the intrinsic ellipticity of the PSF, i.e.
\begin{eqnarray}
 \langle \delta e_{1}\rangle \simeq \frac{e_1}{|e_1|}\sigma(e_1),\\
 \langle \delta e_{2}\rangle \simeq \frac{e_2}{|e_2|}\sigma(e_2).
\end{eqnarray}
The simulated measurements fall very close to the predictions in all panels of the right column of Fig.~\ref{fig:gauss_errstd}, i.e. the predictions are validated by simple Gaussian simulations. 

The slight departure from linearity, for very large binary star fractions, is due to a changing value of $\langle r^2 f/ (1+ f)^2\rangle$, i.e. for large binary fractions the stellar population is modified with respect to populations with a lower fraction of binaries.

The simple linear relation between the bias and  intrinsic ellipticity of the PSF hints at a possible calibration to reduce the effect of binary stars on the PSF shape or, at the very least, on a reliable control of the bias. However, that the lack of knowledge of the distribution of the binary star fraction ($\langle r^2 f/ (1+ f)^2\rangle$) in each specific field used to reconstruct the PSF may complicate the task in practice. 
%
\subsection{Synthetic space-based PSF}
Up to now, we have only considered Gaussian PSFs. In order to test the effect of a more complex PSF, we 
use the {\tt GalSim} software to produce a synthetic PSF similar to future space-based telescopes. More specifically, our PSF is the same size as the Euclid VIS PSF and has six struts. No PSF aberration such as coma or trefoil is applied. The strut angle, with an arbitrarily chosen orientation, is kept constant. 

Similar to the simulations in Sect.~\ref{binfrac}, we build three PSFs with the characteristics shown in Table~\ref{tab:PSF}. We choose the strut-rotated validation case to have the same $e_1$ and $e_2$ input as the PSF along $e_1$.

The results show the same behaviour as for the Gaussian toy model. No significant departure from the predictions are detected. This is likely because we consider here very narrow angular separation binaries, with a companion star well within the central pixel of the PSF. It is therefore not surprising that the main change in shape, due to the binarity, occurs in the very centre of the PSF. 
The details of the PSF shape on larger scales do not matter much. This suggests that (i) simple Gaussian models are sufficient to assess the impact of binarity of the PSF shape, and (ii) our analytical predictions based on Gaussian PSFs are a good approximation of the change in the PSF shape. These predictions can therefore be safely used to predict the PSF reconstruction error as a function of the binary fraction and mean angular separation between the stars and their companions.
%
\subsection{Effect of the number of stars $N_\star$ in the stack}
\label{nstar}
As seen in Sect.~\ref{binfrac} and illustrated in Fig.~\ref{fig:gauss_errstd}, varying the number of stars in the stack $N_\star$ used to reconstruct the PSF changes the reconstruction error. 
Increasing the number of stars increases the sampling of the coordinates for the companion stars. Eventually, for very large values of $N_\star$, the predicted errors as computed in Sect.~\ref{sec:th} are realised by the simulations.

For ${N_\star}=40$ stars, the reconstruction error on the size, as well as the values for the biases on the size and ellipticity, are compatible with the predictions, as in Fig.~\ref{fig:gauss_errstd}. However, the errors on the ellipticity components require larger values for $N_\star$, which also depends on the intrinsic ellipticity of the PSF.
 
This is tested in Fig.~\ref{fig:proba_std} for $N_\star=40, 750, 2100, 10~000$ and for two PSF ellipticities. For low values of $N_\star$, the error is large and independent of the ellipticity of the PSF and the error is dominated by the shot noise due to the number of companions. 
As can be expected, it takes a much larger $N_\star$ to converge to the theoretical predictions, and the smaller the ellipticity, the larger $N_\star$ needed. For example, $N_\star=10~000$ stars in the stack are barely sufficient to converge to the theoretical estimate of the error for PSF2, which has a small $e_1$ component of $-0.021$. 

Figure~\ref{fig:proba_std} shows that when $N_\star$ is very large, the theoretical values for the error estimates on the ellipticity can be modelled by simulating binary populations and the equations in Sect.~\ref{sec:th}. 
However, the figure also shows that if for any reason the prediction we make in Sect.~\ref{sec:th} with Gaussian PSF does not hold true, then the simulations required to replace the theoretical estimates  involve tens of thousands of companions. Given that we also need large numbers of realisations $N_s$, estimating  the effect of binary stars on the PSF reconstruction may quickly become computationally expensive. 

%
\subsection{Dependence on the apparent magnitude of the stars}
\begin{figure}[h]
 \centering
  \includegraphics[width=1\linewidth]{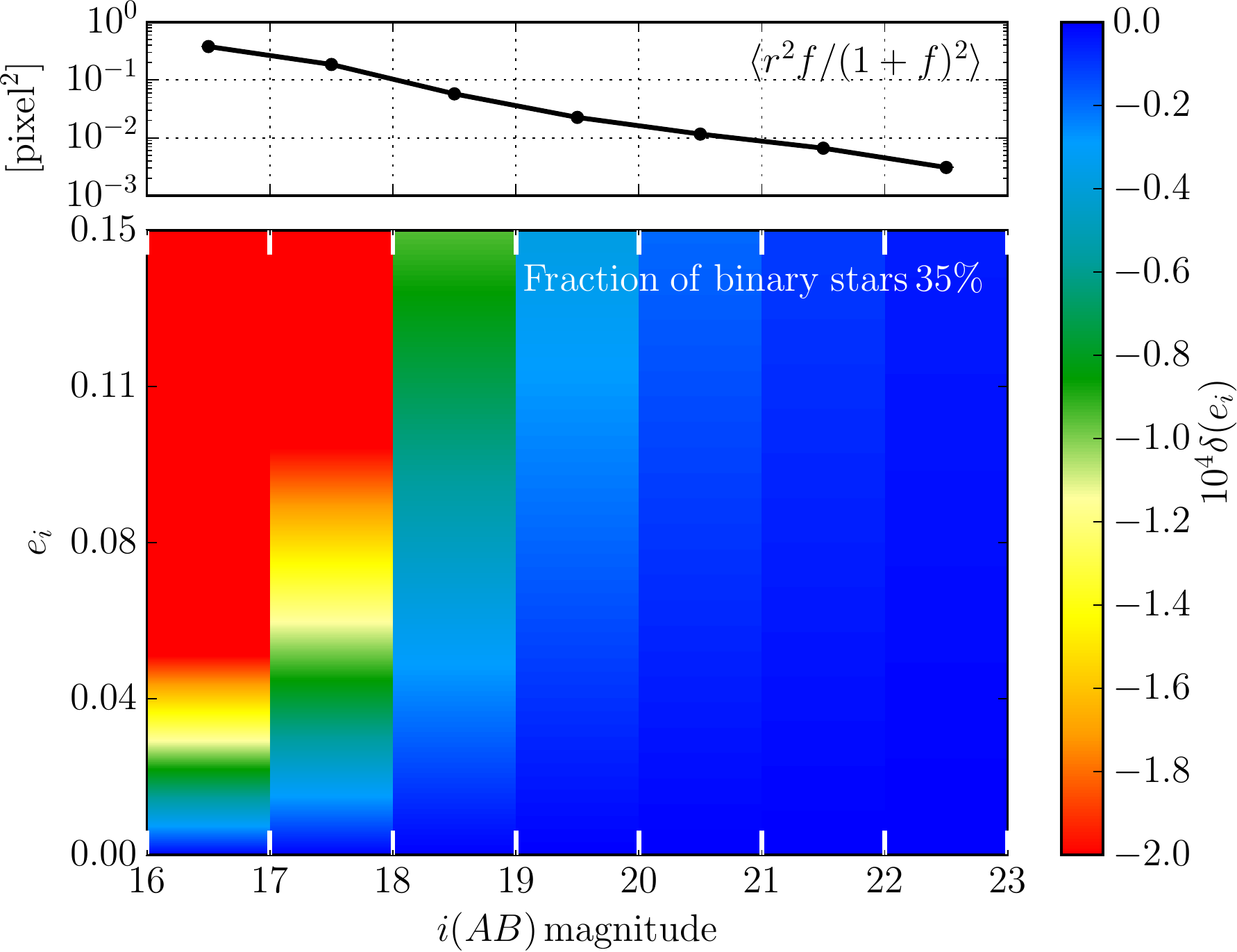}
  \includegraphics[width=1\linewidth]{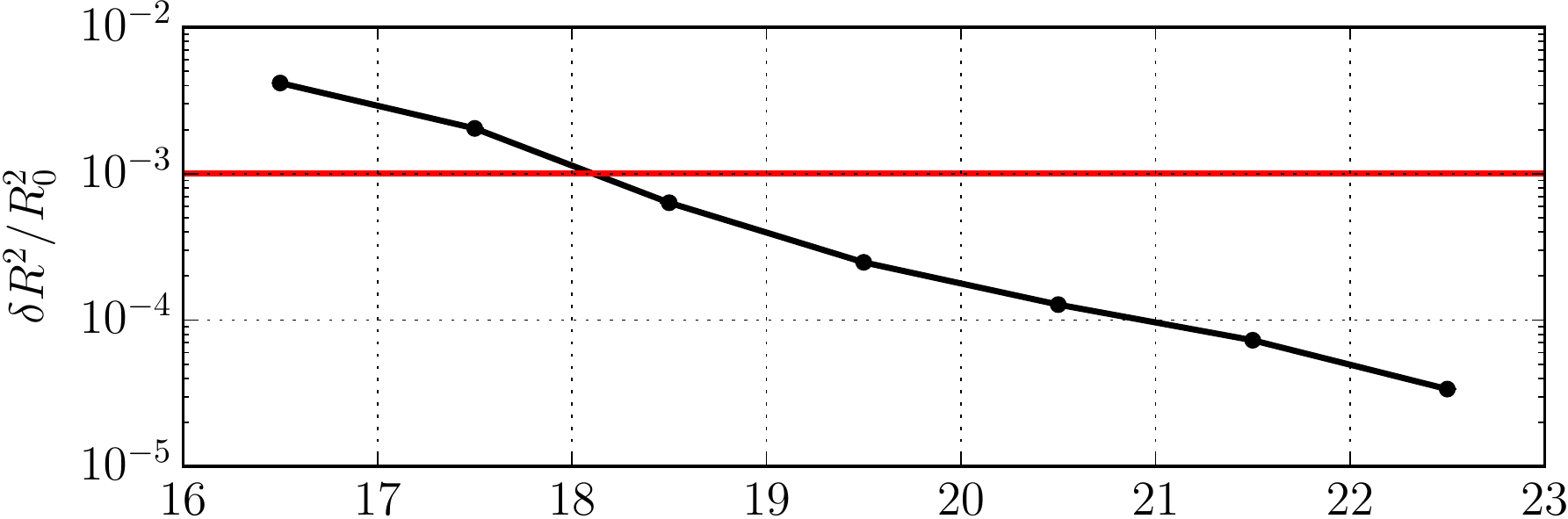}
  \caption{\label{fig:depmag}
  Theoretical reconstruction error according to the formalism developed in Sect.~\ref{sec:th}. The stellar distribution is drawn from realisations of the BMG for pointing coordinates to $\ell=180^\circ,\,b=45^\circ$. \emph{Top:} value of the $\langle r^2 f/ (1+ f)^2\rangle$ parameter as a function of apparent $i(AB)$ magnitude. 
  \emph{Middle:} reconstruction error as a function of the apparent magnitude $i(AB)$ of the stars in the stack and as a function of the complex ellipticity components $e_1$ or $e_2$ (colour code).
  \emph{Bottom:} reconstruction error on the PSF size. The red line indicates typical requirements for stage IV experiments.
  }
\end{figure}
Our ability to resolve a binary star system depends mostly on its physical distance to the Earth, i.e. on its apparent magnitude. The PSF reconstruction errors due to the blend of binaries are correlated with the distance to the binary system through the $\langle r^2 f / (1 + f)^2 \rangle$ parameter. We now turn to the impact of stars in different magnitude bins on the reconstruction of the PSF. 
In doing so, we adopt a binary star fraction of 35\%, which is representative of the mean binary star fraction on the sky \citepalias{Duchene2013}. The resulting reconstruction errors as a function of the $i(AB)$ magnitude are shown on Fig. \ref{fig:depmag}. 
In the figure, we also indicate, as a reference, the total reconstruction error budget for a stage IV weak lensing experiment: $\sigma(e_i) \leq 2\times 10^{-4}$ for the ellipticity components and $\sigma(R^2)/\langle R^2\rangle \leq 1\times10^{-3}$ for the size \citep{Paulin-Henriksson2008}. These values are the total error budgets allowed for these parameters. 

For the  brightest stars with $16 < i(AB)<18$, the error in reconstruction takes the full error budget both on the size and  ellipticity, even for small intrinsic PSF ellipticities. 
Stars in the magnitude range $18<i(AB)<19$ still suffer from a relatively large blending bias, taking up to 60\% of the total error budget on the size and ellipticity.  
Requiring stars fainter than magnitude 20 lowers the blending bias to $\lesssim 10\%$ of the error budget, which is still large enough to be accounted for.

Stars brighter than roughly $i(AB)\simeq 19$ may not be the best choice for PSF calibration strategies, as brighter stars are much more likely to alter the PSF both in size and ellipticity. 
Instead, longer observations of faint and most distant stars are  preferred and  minimise the effect of binary stars. 
Finally, we emphasise that the impact of binary stars on the PSF shape becomes more important with higher spatial resolution if the stellar population remains the same.
Surveys like WFIRST may therefore be more impacted by binary stars than the Euclid-like telescope used in this work. On the other hand, because WFIRST has deeper limiting magnitudes, more faint stars can potentially be used to reconstruct the PSF. 

%
\section{Conclusions} 
\label{sec:conclusions}
Multiple stellar systems are ubiquitous in the Milky Way, with a total fraction of multiple stars of  about 35\% \citepalias{Duchene2013}. These multiple stars, especially when they are not resolved by the instrument, may significantly impact  the PSF measurement given the strong requirements on the modelling of the PSF.
 
In this paper, we present analytical predictions for the PSF reconstruction error introduced by binary stars based on stellar multiplicity parameters and on the intrinsic PSF size and ellipticity. We then verify our predictions by means of realistic numerical simulations.
To do so, we use the BMG \citep{Robin2004} to generate mock stellar distributions. We then add companion stars to produce binary systems following the most recent observational data in \citetalias{Duchene2013}. 
We produce noise-free images used to measure the error and  bias on the PSF ellipticity and size. 

In our simulations, the parameters of the detector and of the PSF are taken to be Euclid-like, but the PSF profile is Gaussian. We also check that our results are valid for more realistic diffraction limited PSFs. 
Binary stars with a separation larger than $r=0.05$\arcsec\ are assumed to be easily identified as multiple and eliminated from the star catalogue to build the PSF. In other words, we only consider double stars that cannot be identified as such but that statistically affect the PSF reconstruction. Our main findings can be summarised as follows.

\begin{itemize}
 \item Binary stars can significantly alter the reconstruction of the PSF. They introduce both a bias and an error on the size and ellipticity of the reconstructed PSF.
 \item The PSF binary reconstruction errors can be predicted analytically as a function of characteristics of the stellar population considered, i.e. the separations and intensity ratio of the stars, and as a function of the intrinsic PSF size and ellipticity.
 \item The effect of the binaries cannot be suppressed by averaging a large number of stars.
 \item The analytical predictions are supported by numerical experiments. The PSF used in the simulations are Gaussian and diffraction-limited.
 \item Bright stars of magnitudes $i(AB)\lesssim 19$ are most affected by binarity. These stars are either massive stars with very likely companions or they are nearby stars, with potentially large angular separations to their companion(s).
 \item Fainter stars in the range $18 < i(AB) < 19$) may cause signi\-ficant biases on the PSF shape. If no mitigation scheme is found, these stars can take on their own up to $\sim60\%$ of the total budgeted error on the size and ellipticity described in 
 \citet[]{Paulin-Henriksson2008} and \citet[]{Cropper2013}.
 \item Blending errors in systems fainter $20 < i(AB) < 21$ can still contribute up to 10\% to the typical error budget on the size and ellipticity.
 \item If any PSF calibration fields are chosen for space missions, targeting fields of faint stars can be preferable to bright stars to avoid nearby binary systems that have larger effects on the PSF shape.
 \item The total fraction of binaries varies with Galactic longitude. There are hints that the binary star fraction increases with Galactic latitude. Thus, the variations in the underlying stellar populations can have a varying effect on the PSF shape as a function of Galactic coordinates. This may also imply  a bias on the shear power spectrum that must be controlled.
\item Different stellar types differ in their fraction of multiple systems and also in the distribution of the parameters of the orbits. Regions of the sky with lower binaries fractions could be exploited to quantify the effect of binary stars on the PSF determination differentially across the sky.
\end{itemize}

Given the above conclusions, it will be important that future space-based weak lensing surveys account for the effect of binary stars on the PSF determination. In particular, their contribution to the error budget on the PSF shape (both the error and the bias) must be estimated and any PSF reconstruction method must account for the extra error caused by binarity. The above conclusions depend on the ability to clean star samples from multiple systems. This can be done in many ways and with different efficiency depending on signal-to-noise ratio and stellar type. Gaia may provide the spectral type of the stars and, in some cases, identify binary stars from their apparent motion on the plane of the sky. Alternatively, specific image processing techniques may be used, but these remain to be devised. 

\begin{acknowledgements}
We are indebted to the anonymous referee whose valuable comments improved the quality of this work.  
The authors would like to thank Laurent Eyer, David Harvey, Henk Hoekstra, Matthew Nichols, Pierre North, and Thomas Kitching for useful discussions.
This work is supported by the Swiss National Science Foundation (SNSF). \end{acknowledgements}


\bibliographystyle{aa} 
\bibliography{bib} 

\end{document}